%% file: main.tex
\documentclass[twocolumn]{autart}    

\usepackage{graphicx}          
\usepackage{amsmath} 
\usepackage{amssymb}  
\usepackage{algorithm}
\usepackage[noend]{algpseudocode}
\usepackage{color}
\usepackage{graphicx}
\usepackage{booktabs}

\begin{document}

\begin{frontmatter}

\title{State Constrained Stochastic Optimal Control for Continuous and Hybrid Dynamical Systems Using DFBSDE\thanksref{footnoteinfo}}

\thanks[footnoteinfo]{This work is supported in part by the NSF under grant DMS-1907518. An earlier version~\cite{9482832} of a portion of this paper was presented at the 2021 American Control Conference (Virtual), May 2021.}

\author[NYU]{Bolun Dai}\ead{bd1555@nyu.edu}, 
\author[NYU]{Prashanth Krishnamurthy}\ead{prashanth.krishnamurthy@nyu.edu},
\author[NCSU]{Andrew Papanicolaou}\ead{apapani@ncsu.edu},
\author[NYU]{Farshad Khorrami}\ead{khorrami@nyu.edu}
\address[NYU]{New York University, Brooklyn, NY, 11201}
\address[NCSU]{North Carolina State University, Raleigh, NC, 27695}

\begin{keyword}                           
Stochastic control, Optimal control, Forward and backward stochastic differential equations         
\end{keyword}                             

\input{abstract}

\end{frontmatter}

\newcommand{\x}{\mathbf{x}}
\newcommand{\y}{\mathbf{y}}
\newcommand{\bu}{\mathbf{u}}
\newcommand{\w}{\mathbf{w}}
\newcommand{\F}{\mathbf{F}}
\newcommand{\G}{\mathbf{G}}
\newcommand{\U}{\mathbf{U}}
\newcommand{\R}{\mathbf{R}}
\newcommand{\V}{\mathbf{V}}

\input{intro.tex}
\input{formulation.tex}
\input{dfbsde.tex}
\input{simulation.tex}
\input{conclusion.tex}

\bibliographystyle{plain}        
\bibliography{autosam.bib}           
\end{document}

%% file: abstract.tex
\begin{abstract} 
We develop a computationally efficient learning-based forward-backward stochastic differential equations (FBSDE) controller for both continuous and hybrid dynamical (HD) systems subject to stochastic noise and state constraints. Solutions to stochastic optimal control (SOC) problems satisfy the Hamilton–Jacobi–Bellman (HJB) equation. Using current FBSDE-based solutions, the optimal control can be obtained from the HJB equations using deep neural networks (e.g., long short-term memory (LSTM) networks). To ensure the learned controller respects the constraint boundaries, we enforce the state constraints using a soft penalty function. In addition to previous works, we adapt the deep FBSDE (DFBSDE) control framework to handle HD systems consisting of continuous dynamics and a deterministic discrete state change. We demonstrate our proposed algorithm in simulation on a continuous nonlinear system (cart-pole) and a hybrid nonlinear system (five-link biped).
\end{abstract}

%% file: intro.tex
\section{Introduction}
Optimal control has been applied in various applications, e.g., robotic control~\cite{DBLP:conf/icra/MistryBS10}~\cite{DBLP:journals/corr/abs-2205-05429} and navigation~\cite{DEMETRIOU2020108859}. Optimal control problems are often cast as the minimization of a cost function, which the solution can be found via numerical optimization techniques~\cite{diehl2009efficient}~\cite{DaiHKK23}~\cite{DaiKKGTK23}. Recently, optimal control has been used to solve increasingly complex control problems~\cite{neunert2018whole} thanks to the enhanced computational abilities and optimization software and the adoption of neural networks~\cite{DBLP:conf/iclr/ChenSZ19}. Applying optimal control requires accurate modeling of the system of interest. However, unmodeled processes are common in real-world systems, e.g., measurement noise and external forces with unknown distributions~\cite{lorenzen2017stochastic}~\cite{hewing2019scenario}. We adopt SOC~\cite{crespo2003stochastic} techniques to consider the uncontrolled inputs explicitly.

Recently, neural networks have been widely used to solve SOC problems~\cite{PersioG21}~\cite{DBLP:journals/corr/abs-2107-07931}. A promising direction for deep neural network (DNN) based SOC solutions is formulating the problem as an FBSDE~\cite{DBLP:conf/rss/WangPT19}. FBSDE enables the numerical estimation of the solution to the HJB equation, i.e., the value function with respect to a cost function. The optimal control at each state then can be analytically obtained using the estimated value function. The introduction of deep neural networks~\cite{DBLP:conf/rss/WangPT19} to FBSDEs improves the value function estimation. Compared to nonlinear model predictive control (MPC) approaches, DFBSDE-based approaches are computationally more efficient during inference by moving the heavy computation offline. On top of the vanilla DFBSDE, work has been done in introducing state constraints using penalty functions~\cite{9482832} and appending a differentiable quadratic program~\cite{Pereira0ET20}. In this work, we explore a penalty function based approach since it is faster in training and inference.

Another key aspect of this paper is the control of HD systems. The type of HD systems considered consists of two phases: first, a phase of continuously evolving dynamics, and second, a phase of a discrete jump. Such HD systems are very common, e.g., walking robots~\cite{GRIZZLE20141955}. Thus, it is worthwhile to understand how it is typically controlled. The control of walking robots is traditionally achieved with a hierarchical control architecture. A reduced-order model is used for motion generation. The generated motion is tracked via a tracking controller~\cite{DBLP:conf/icra/MistryBS10}. Another approach is to see the HD system as a combination of multiple phases of continuous dynamics~\cite{9562022}. Using this approach, the controller only needs to work well for the continuous dynamics and be robust to changes caused by discrete jumps. This formulation is end-to-end, which better suits DFBSDE.

This paper proposes a SOC setting for nonlinear systems that incorporates state constraints and adapts the method to handle HD systems. The main contribution of this paper is threefold: (1) proposed the state constraint DFBSDE algorithm and adapted it to handle HD systems; (2) devised a new training loss and controller ensemble setting for high-dimensional HD systems; (3) provided simulation studies on both continuous and HD systems to show the efficacy of our approach.  In relationship with our previous work~\cite{9482832}, this paper provides a more detailed development of the methodology, extension to hybrid systems, and application to bipeds. The remainder of this paper is structured as follows. In Section II, the state-constrained SOC formulation is given. Section III presents a derivation connecting the SOC formulation and FBSDEs. Also, Section III presents the incorporation of control saturation and state constraints. Then using the derived formulation, a deep neural network-based algorithm is presented to solve the FBSDE with state constraints and control saturation. The last part of Section III discusses adapting the DFBSDE algorithm to HD systems. In Section IV, simulation studies are presented for a continuous dynamical system, namely the cart-pole, and a five-link biped~\cite{DBLP:journals/siamrev/Kelly17}, which is HD. 

%% file: formulation.tex
\section{Problem Formulation}
In this section, we outline the SOC problem under state constraints. We consider a filtered probability space $(\Omega, \mathcal{F}, \{\mathcal{F}_{t}\}_{t\geq0}, \mathbb{P})$, where $\Omega$ is the sample space, $\mathcal{F}$ is the $\sigma$-algebra over $\Omega$, $\mathbb{P}$ is a probability measure, and $\{\mathcal{F}_{t}\}_{t\geq0}$ is a filtration with index $t$ denoting time. A controlled affine system with noise represented by stochastic processes can be described using a stochastic differential equation (SDE)
\begin{equation}
	\label{eq:SDE2}
	d{\x}(t) = (\F(\x(t)) + \G(\x(t))\bu)dt + \Sigma(\x(t))d{\w}(t)
\end{equation}
with initial state $\x_0\in\mathbb{R}^n$ and $\w(t)\in\mathbb{R}^\nu$ an $\mathcal{F}_{t}$-adapted Brownian motion. Throughout this paper, we will write $\x(t)$ and other stochastic processes with argument $t$ omitted whenever appropriate. The states are denoted by $\x\in\mathbb{R}^n$ and the control input by $\bu\in\mathbb{R}^m$. In \eqref{eq:SDE2}, $\F: \mathbb{R}^n\rightarrow\mathbb{R}^n$ represents the drift, $\G: \mathbb{R}^n\rightarrow\mathbb{R}^{n\times m}$ represents the control influence, and $\Sigma: \mathbb{R}^n\rightarrow\mathbb{R}^{n\times\nu}$ represents the diffusion (influence of the Brownian motion to the state). We assume that $\mathrm{range}(\G) \subseteq \mathrm{range}(\Sigma)$, i.e., the noise enters wherever the control input appears (in addition to possibly elsewhere). The state constrained SOC problem is to find a controller $\bu(\x)$ that minimizes an objective function $J^\bu(\x, t)\in\mathbb{R}_+$ under a set of state constraints. The objective function is denoted as
\begin{align}
    J^\bu(\x, t) =&\ \mathbb{E}\Big[\mathbf{q}_N(\x(T)) + \int_{t}^{T}\Big(\mathbf{q}(\x) + \mathbf{r}(\bu)\Big)d\tau\Big]
    \label{eq:objective_optimization}
\end{align}
where $\mathbb{E}$ represents expectation, $T\in\mathbb{R}_+$ is the terminal time, $\mathbf{q}_N:\mathbb{R}^n\rightarrow\mathbb{R}_+$ is the terminal state cost, $\mathbf{q}:\mathbb{R}^n\rightarrow\mathbb{R}_+$ is the instantaneous state cost, and the instantaneous control cost is $\mathbf{r}:\mathbb{R}^m\rightarrow\mathbb{R}_+$. The controller $\bu(\x(t))$ is a function of the state, this dependency will be omitted in the remainder of this paper for brevity, and the controller will be written as $\bu$. Without loss of generality, we consider state constraints in the following form $\mathbf{c}(\x) \leq \mathbf{b}$, where $\mathbf{c}(\x)\in\mathbb{R}^r$ is a vector of functions of the state, and $\mathbf{b}\in\mathbb{R}^r$ represents the element-wise upper bound of $\mathbf{c}(\x)$.  Control saturation (with $\U_{\max}\in\mathbb{R}_+^m$) can also be introduced into the SOC formulation
\begin{equation}
    \bu\in\mathcal{U} = \{\bu\mid|\bu_i|\leq \U_{i, \max},\ i = 1, \hdots, m\}
    \label{eq:control_saturation}
\end{equation}
where $\bu_i$ and $\U_{i, \max}$ are the $i^{th}$ element of $\bu$ and $\U_{\max}$, respectively. Note that all operations with stochastic processes are understood a.s. (almost surely, i.e., with probability 1) unless specifically indicated. In summary, we have the SOC problem as
\begin{align}
\label{eq:soc_problem}
    \min_{\bu\in\mathcal{U}}\ &\ J^\bu(\x_0, t_0)\ \mathrm{subject\ to}\ \ \eqref{eq:SDE2}\ \mathrm{and}\ \mathbf{c}(\x) \leq \mathbf{b}\ \mathrm{a.s.}
\end{align}

%% file: dfbsde.tex
\section{Method}
In this section, we first show how~\eqref{eq:soc_problem} can be written as an FBSDE. Then, state constraints are considered. Subsequently, adaptations for HD systems are presented. Finally, a DNN-based solution to the FBSDE is proposed.

\subsection{SOC to FBSDE}
The derivation presented in this section was originally proposed in~\cite{DBLP:conf/rss/WangPT19}, it is summarized here for completeness. Let $\V(\cdot, \cdot)$ be the value function $\V(\x, t)\in\mathbb{R}_+$ defined as
\begin{equation}
    \V(\x, t) := \inf_{\bu\in\mathcal{U}} J^{\bu}(\x, t).
\end{equation}
Using~\eqref{eq:objective_optimization} and Bellman's principle~\cite{Bellman03} yields
\begin{subequations}
\begin{align}
    & \V_t(\x, t) + \mathcal{L}\V(\x, t) + \mathbf{h}(\x, \V_\x(\x, t)) = 0\\
    & \V(\x(T), T) = \mathbf{q}_N(\x(T))
\end{align}
\label{eq:HJB}%
\end{subequations}%
where $\mathbf{h}$ denotes the Hamiltonian
\begin{equation}
    \mathbf{h}(\x, \V_\x) = \inf_{\bu\in\mathcal{U}}\Big(\mathbf{q}(\x) + (\G(\x)\bu)^T\V_\x+\mathbf{r}(\bu)\Big),
    \label{eq:hamiltonian}
\end{equation}
the differential operator of $\V$ is given by
\begin{equation}
    \mathcal{L}\V(\x, t) = \frac{1}{2}\mathrm{trace}\Big(\Sigma\Sigma^T\V_{\x\x}(\x, t)\Big) + \F^T(\x)\V_\x(\x, t),
    \label{eq:generator_function}
\end{equation}
$\V_t(\x, t)\in\mathbb{R}$ is the partial derivative of $\V$ with respect to $t$, and $\V_{\x}(\x, t)\in\mathbb{R}^{n\times1}$ and $\V_{\x\x}(\x, t)\in\mathbb{R}^{n\times n}$ denote the first and second partial derivatives, respectively, of $\V$ with respect to $\x$.
To consider control constraints~\cite{DBLP:conf/rss/WangPT19}, we define $\mathbf{r}(\bu)$ as
\begin{equation}
\label{eq:control_constraint_S}
    \mathbf{r}(\bu) = \sum_{i=1}^{m}\mathbf{S}_i(\bu_i) = \sum_{i=1}^{m}c_i\int_{0}^{\bu_i}{\mathrm{sig}^{-1}\Big(\frac{v}{\bu_i^{\max}}\Big)dv}
\end{equation}
with $\bu_i^{\max}$ being the $i$-th element of $\mathbf{U}_{\max}$ and
\begin{equation}
    \mathrm{sig}(v) = \frac{2}{1 + e^{-v}} - 1.
    \label{eq:sig_func}
\end{equation}
In~\eqref{eq:control_constraint_S}, $c_i$ weights the importance between different control inputs. Using first-order conditions to solve for an analytic solution of the optimal control action that achieves the infimum of the Hamiltonian yields
\begin{equation}
    \G^T(\x)\V_\x + \mathbf{R}\mathrm{sig}^{-1}(\frac{\bu^*}{\mathbf{U}_{\mathrm{max}}}) = 0,
    \label{eq:u_star_eq}
\end{equation}
with $\R = \mathrm{diag}(c_1, \cdots, c_m)$. Solving~\eqref{eq:u_star_eq} for $\bu^*$ in yields
\begin{equation}
    {\bu}^*(\x, t) = \mathbf{U}_{\max}*\mathrm{sig}(-\R^{-1}\G^T(\x)\V_\x(\x, t))
    \label{eq:control_constraint}
\end{equation}
with ``$*$" denoting element-wise multiplication. The control is saturated between $[-\mathbf{U}_{\max}, \mathbf{U}_{\max}]$. Define $\y(t) = \V(\x(t), t)$. From~\eqref{eq:hamiltonian}, we can write a stochastic system for the optimal value function
\begin{subequations}
\label{eq:FBSDE_undo}
\begin{align}
\label{eq:FBSDE_b1}
-d{\y}(t) &= (\mathbf{q}(\x) + \mathbf{r}(\bu^*))dt - \V_\x^T(\x, t)\Sigma^T(\x)d{\w}\\
\label{eq:FBSDE_f1}
d{\x}(t) &= (\F(\x) + \G(\x)\bu^*)dt + \Sigma(\x)d{\w}\\
\label{eq:FBSDE_b2}
\y(T) &= \mathbf{q}_N(\x(T))\\
\label{eq:FBSDE_f2}
\x(0) &= \x_0
\end{align}
\end{subequations}
which is in the form of a FBSDE (the forward part consists of~\eqref{eq:FBSDE_f1}~\eqref{eq:FBSDE_f2}, the backward part consists of~\eqref{eq:FBSDE_b1}~\eqref{eq:FBSDE_b2}). The forward and backward parts are of the form of a forward SDE (FSDE) and a backward SDE (BSDE), respectively. For details regarding the derivation from~\eqref{eq:hamiltonian} to~\eqref{eq:FBSDE_undo}, the reader is referred to~\cite{DBLP:conf/rss/WangPT19}\cite{9482832}.

\subsection{Handling State Constraints}
\begin{figure}[t!]
    \centering
    \includegraphics[width=0.49\textwidth]{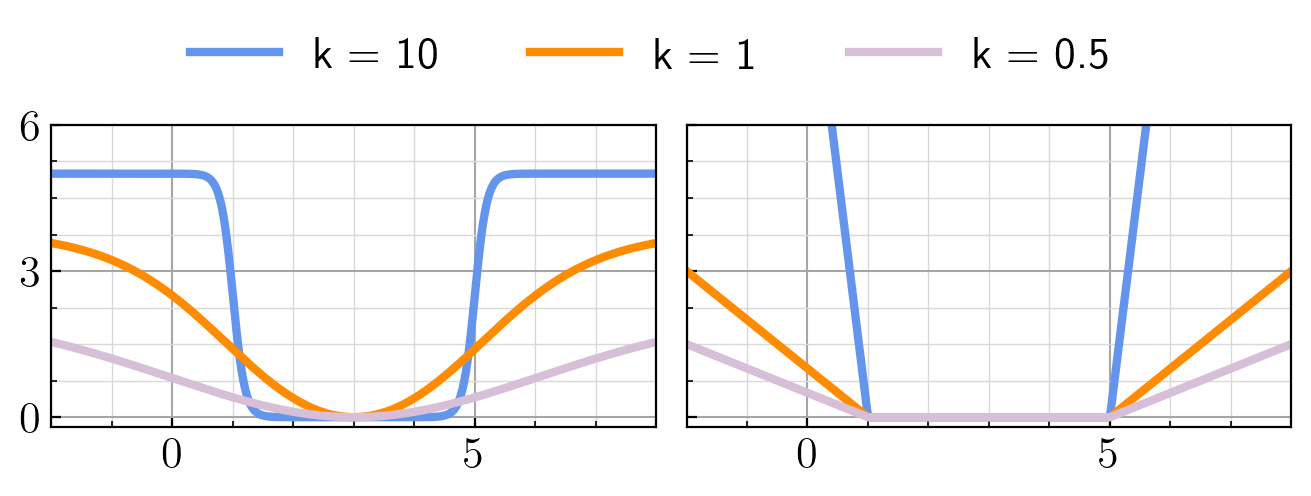}
    \caption{The left figure shows $\mathbf{p}(\x)$ in~\eqref{eq:soft_constraint} under different values of $k$, with $\mu = 3$, $L = 5$, $\mathbf{c}(\x) = \x$, $\mathbf{b}_{\min} = 1$, $\mathbf{b}_{\max} = 5$ and $\x$ being a scalar. The right figure shows $\mathbf{p}(\x)$ in~\eqref{eq:penalty_relu} under different values of $k$, with the same parameters.}
    \label{fig:soft_constraints}
\end{figure}
In this section, we consider incorporating state constraints into the FBSDE framework using a penalty function based approach. A perfect penalty function would be zero inside the constraint boundary and infinity outside. We propose two differentiable and numerically stable alternatives to a perfect penalty function. The first option is a penalty function based on logistic functions (PFL). When the state constraint has both an upper bound $\mathbf{b}_{\max}$ and lower bound $\mathbf{b}_{\min}$, the $i$-th element of PFL would be
\allowdisplaybreaks
\begingroup
\begin{align}
\label{eq:soft_constraint}
    \mathbf{p}_i(\x) =&\ \frac{L}{1 + e^{-\mathbf{k}(\mathbf{c}_i(\x) - \mathbf{b}_{i, \max})}} - \frac{L}{1 + e^{-\mathbf{k}(\mathbf{c}_i(\x) - \mathbf{b}_{i, \min})}}\nonumber\\
          &+ L - \frac{2L}{1 + e^{-\mathbf{k}(\mu_i - \mathbf{b}_{i, \max})}}
\end{align}
\endgroup
where $L\in\mathbb{R}^+$ determines the maximum value of the penalty, $\mathbf{k}\in\mathbb{R}^+$ determines the steepness of the boundary (larger $k$ leads to steeper boundaries), and $\mu = (\mathbf{b}_{\min} + \mathbf{b}_{\max})/2$. The proposed penalty function consists of two parts: the first consists of two logistics functions, which give the valley shape; the second sets the minimum value of the penalty function to zero. The second option is a rectified-linear-unit (ReLU) based function which is defined as $\mathbf{ReLU}(x) = \max(x, 0)$. When an upper and lower bound exist, the penalty function is
\begin{align}
    \label{eq:penalty_relu}
    \mathbf{p}_i(\x) =&\ \mathbf{k}(\mathbf{ReLU}(\mathbf{b}_{i, \min} - \mathbf{c}_i(\x))\nonumber\\
    &+ \mathbf{ReLU}(\mathbf{c}_i(\x) - \mathbf{b}_{i, \max})).
\end{align}
Similar to~\eqref{eq:soft_constraint}, $\mathbf{k}$ represents steepness of the constraint boundaries. However, in~\eqref{eq:penalty_relu}, there is no bound on the penalty function magnitude. An example of these two types of penalty functions with varying steepness is shown in Fig.~\ref{fig:soft_constraints}. When only an upper bound exists, i.e., $\mathbf{c}(\x) \leq \mathbf{b}$, the lower bound is set to be $-\infty$, and when only a lower bound exists, i.e., $\mathbf{c}(\x) \geq \mathbf{b}$, the upper bound is set to be $\infty$. The main difference between these two penalty functions is their values within the constraint boundary. For PFL, the value of the penalty function is only close to zero when a large $\mathbf{k}$ value is utilized. Otherwise, states within but close to constraint boundaries will have a positive penalty function value. The ReLU-based penalty function is always zero within the bound. Specific use cases are demonstrated in Section~\ref{sec:simulations}. After applying penalty function $\mathbf{p}(\x)$, the instantaneous state cost is $\bar{\mathbf{q}}(\x) = \mathbf{q}(\x) + \mathbf{p}(\x)$. Ideally, we would pick larger $\mathbf{k}$ and $L$ to ensure a steeper boundary and larger penalty. While a controller without access to measurements of future noise can not {\em guarantee} that state constraints will be met for all time due to stochastic nature of the system dynamics, note that the penalty functions guide the learning towards a robust controller that does not violate state constraints (at least under the disturbances seen during training). The expected value of the integral of $\mathbf{p}(\x)$ is at most $J^\mathbf{u}$. If $J^\mathbf{u}$ is finite and $\mathbf{p}(\x)$ is a perfect penalty function, the constraint is satisfied except on a set of measure zero.

\subsection{Deep FBSDE Algorithm}
\begin{figure}[t!]
    \centering
    \includegraphics[width=0.49\textwidth]{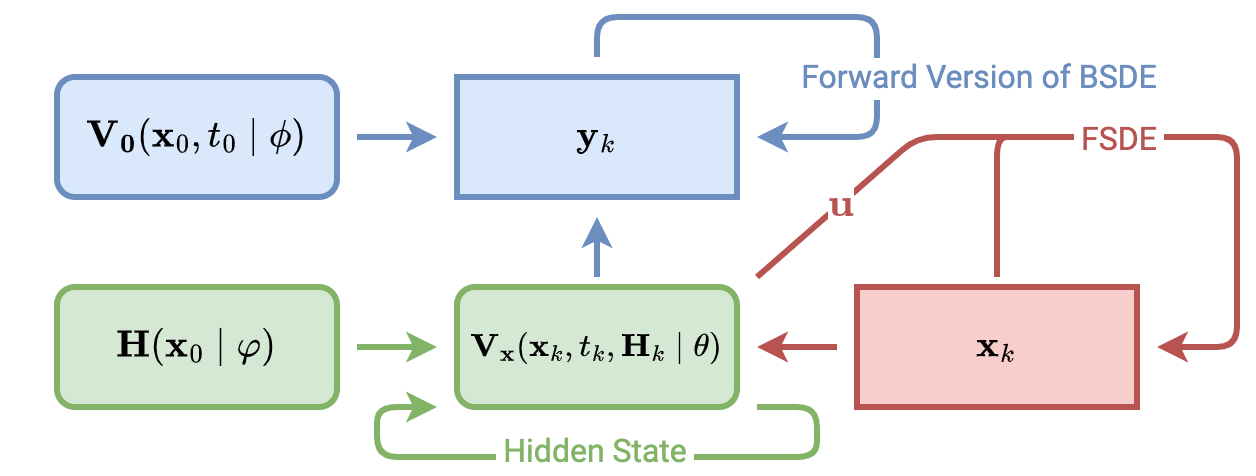}
    \caption{The DFBSDE architecture for one time step. Curve-edged boxes contains DNNs, and sharp-edged boxes represent intermediate calculations. Everything in green contains the LSTM network, red and blue colors represent the FSDE and the forward version of the BSDE, respectively.}
    \label{fig:FBSDEModel}
\end{figure}
The solution of the FBSDE in~\eqref{eq:FBSDE_undo} requires backward integration in time and the fact that $\V_\x(\x)$ is unknown creates additional difficulties in utilizing classical numerical integration approaches.
To deal with backward integration in time, following the formulation in~\cite{DBLP:conf/rss/WangPT19}, we can estimate the value function at time zero using a DNN, which is parameterized by $\phi$, i.e., $\V_\mathbf{0}(\x, t_0)$. This allows the forward integration of the BSDE. We can then rewrite the FBSDE as two FSDEs. For unknown $\V_\x(\x, t)$ values, it can also be estimated using a DNN parameterized by $\theta$, i.e., $\V_\x(\x, t\mid\theta)$. We deploy an LSTM-based architecture~\cite{HochreiterS97} for $\V_\x(\x, t\mid\theta)$, which has been shown to be superior to dense-layer-based architectures~\cite{DBLP:conf/rss/WangPT19}. Linear regression based approaches have also been considered, but  are inferior due to compounding error. Assuming the initial state and time horizon are fixed~\cite{DBLP:conf/rss/WangPT19}, the initial value function is a learned fixed value $\V_\mathbf{0}(\phi)$. Due to the use of LSTMs, two changes need to be made to the DFBSDE algorithm. First, a separate network $\mathbf{H}_0(\varphi)$ is used to estimate the initial hidden state values. Second, $\y_k$ will also depend on the hidden state values $\mathbf{H}_{k}$.
The FBSDE can be forward integrated by discretizing the time as follows
\begingroup
\allowdisplaybreaks
\begin{subequations}
\begin{align}
    \mathbf{y}_k &= V_\mathbf{0}(\x_0, t_0\mid\phi) - \sum_{i=1}^{k}(\bar{\mathbf{q}}(\x_{i-1}) + \mathbf{r}(\bu_{i-1}^*))\Delta{t}\nonumber\\
    +& \sum_{i=2}^{k}\V_\x^T(\x_{i-1}, t_{i-1}, \mathbf{H}_{i-1}\mid\theta)\Sigma^T(\x_{i-1})\Delta{\w}_{i-1}\nonumber\\
    +& \V_\x^T(\x_0, t_0, \mathbf{H}_0(\varphi)\mid\theta)\Sigma^T(\x_0)\Delta{\w}_0\\
    \x_k &= \x_0 + \sum_{i=1}^{k}\Delta{\mathbf{x}_{i-1}}
\end{align}
\end{subequations}
\endgroup
where $k\geq1$, $\Delta{t} = T/N$, $\Delta{\mathbf{x}}_k = \dot{\mathbf{x}}(k\Delta{t})\Delta{t}$ and $\Delta{\mathbf{w}}_k = \dot{\mathbf{w}}(k\Delta{t})\Delta{t}$. The calculation of $\mathbf{x}_k$ and $\mathbf{y}_k$ is illustrated in Fig.~\ref{fig:FBSDEModel}. To learn the weights, we minimize the least square loss between the estimated terminal value function and the measured terminal cost,
\begin{equation}
    \mathbf{L}_\mathrm{FBSDE}(\theta, \phi, \varphi) = \sum_{j=1}^{M}\|\mathbf{q}_N(\x_N^j) - \y_N^j\|^2
    \label{eq:fbsde_loss}
\end{equation}
where $M$ is the batch size and the superscript $j$ denotes the $j$-th set of data in the batch. The detailed DFBSDE algorithm is shown in Algorithm~\ref{alg:dfbsde}.

\subsection{Penalty Function Update}
For the penalty function, if $\mathbf{k}$ is set to a large value in the initial stage of training, numerical instabilities can arise due to large gradients generated by states within the steep region near the constraint boundary. On the other hand, if the values of $\mathbf{k}$ are kept small, it would not be effective since the penalty for constraint violation is small. Thus, we propose to update $\mathbf{k}$ during training so that it gradually increases. The update scheme is as follows. For a given $\mathbf{k}$, the DNN is trained until convergence; then, $\mathbf{k}$ is updated. A metric to check for convergence is the variance of the episode-wise cost over a few episodes; the episode window size is denoted as $\eta$. If the iterates are converging, the variance will be small. Convergence is determined by comparing the square root of the variance with a threshold $\beta\in\mathbf{R}_+$; if it is smaller than $\beta$, $\mathbf{k}$ is updated as $\mathbf{k} \leftarrow \mathbf{k} + \delta$. This condition is checked every $\eta$ iterations. Additionally, if the condition is not satisfied after $\eta^\prime$ iterations, then also $\mathbf{k}$ is updated since it could be stuck at a local minimum. Empirically, we have found $\eta = 500$, $\mathbf{k} = 1.5$, $\delta = 0.5$, and $\Delta_\delta = 0.25$ to be reasonable values to start with. Both $\delta$ and $\Delta_\delta$ can be increased if training is stable and faster growth of $\mathbf{k}$ is desired; otherwise, $\delta$ and/or  $\Delta_\delta$ should be reduced. During training, the variance decreases. Thus, $\beta$ should also decrease after each update: $\beta = \gamma\beta$, with $\gamma\in(0, 1)$. To make the decrease in $\beta$ smoother, we gradually increase $\gamma$ to 1 using $\gamma = \gamma + \Delta$ with $\Delta\in\mathbb{R}_+$. Similarly, the acceleration of $k$ is made negative, i.e., $\delta = \delta + \Delta_\delta$, with $\Delta_\delta\in\mathbb{R}_-$, leading to finer-grained changes in $k$ at later stages of training. The initial value of $\beta$ can be determined after the training converges for the first iteration of Algorithm~\ref{alg:k_update}. From empirical evaluations, we find $\gamma = 0.9$ and $\Delta = 0.02$ to be reasonable values. The penalty function update scheme is shown in Algorithm~\ref{alg:k_update}.  

\subsection{Adaptation to Hybrid Dynamics}
\label{sec:hybrid}
This section will discuss the adaptations required for the DFBSDE algorithm to handle HD systems. HD systems consist of two phases: a continuous dynamic phase and a phase of a deterministic discrete jump. Such an HD system is very common, e.g., a biped. The FBSDE formulation in~\eqref{eq:FBSDE_undo} is only for continuous dynamics. Inspired by the cyclic motion in human walking, we treat the hybrid dynamics as having multiple cycles, where the dynamics are continuous within each cycle. For the case of bipedal locomotion, each cycle will be a footstep; at the end of each cycle, the swing foot lands on the ground (more on this in Section~\ref{sec:simulations}). The main challenges in adapting the DFBSDE framework to HD systems are robustness to the initial state and assurance of cyclic motion. This requires the initial value function estimator to be robust to a wide range of states. The loss function~\eqref{eq:fbsde_loss} only trains the estimator for a fixed $\x_0$. To ensure that the learned value function estimator is robust to variations in the initial state, we can train it such that it provides an accurate estimation for all of the states recorded. The measured cost-to-go can approximate the value function 
\begin{align}
    \tilde{\V}(\x_k, t_k) =&\ \mathbf{q}_N(\x_N) + \sum_{i=k}^{N-1}(\bar{\mathbf{q}}(\x_{i}) + \mathbf{r}(\bu_{i}^*))\Delta{t}\nonumber\\
    -&\ \V_\x^T(\x_{i}, t_{i}, \mathbf{H}_{i}\mid\theta)\Sigma^T(\x_{i})\Delta{\w}_{i}.
    \label{eq:value_function_target}
\end{align}
Using~\eqref{eq:value_function_target}, we can learn the initial value function using
\begin{equation}
    \mathbf{L}_\V = \sum_{j=1}^{M}\sum_{i = 0}^{N}\|\V_\mathbf{0}(\x_i^j, t_i\mid\phi) - \tilde{\V}(\x_i^j, t_i)\|^2.
\end{equation}
Thus, the loss function for the Hybrid FBSDE (HFBSDE) becomes
\begin{equation}
    \mathbf{L}_{\mathrm{HFBSDE}} = \mathbf{L}_\mathrm{FBSDE} + \lambda\mathbf{L}_\V
    \label{eq:fbsde_loss_hybrid}
\end{equation}
with the terminal cost calculated using the state after the jump $\x_N^{+}$ and $\lambda\in\mathbb{R}_+$ adjusts the weighting between the two losses. This terminal cost encourages cyclic motion.

\input{dfbsde_alg.tex}
\input{penalty_algorithm}

%% file: dfbsde_alg.tex
\begin{algorithm}[b!]
\caption{State Constrained DFBSDE Controller}
\begin{algorithmic}[1]
    \State Get Initial state and state dynamics, cost function parameters, penalty function, $N$: Time horizon, $N_I$: Number of iterations, $M$: Batch size, $\Delta{t}$: Time step, $\lambda$: Weight decay parameter, and state constraint parameters (refer to Algorithm~\ref{alg:k_update});
    \State Initialize $\theta$, $\phi$, $\varphi$, $k$, $\delta$, $\beta$, $\gamma$;
    \For {$n_I = 1$ to $N_I$, $m = 1$ to $M$, $k = 0$ to $N-1$}
        \State Calculate $t_k = k\Delta{t}$ and $\V_{\x}^m(\x_k, t_k\mid\theta)$;
        \State Calculate optimal control as in~\eqref{eq:control_constraint};
        \State Sample Brownian noise: $\Delta{\w}_k^m\sim\mathcal{N}(0, \Delta{t})$;
        \State Update value function $\mathbf{y}_k^m$ and system state $\mathbf{x}_k^m$;
        \State Compute target terminal cost;
        \State Compute $\mathbf{L}_\mathrm{FBSDE}$~\eqref{eq:fbsde_loss} or $\mathbf{L}_\mathrm{HFBSDE}$~\eqref{eq:fbsde_loss_hybrid}
        \State Update $\theta$, $\phi$ and $\varphi$ and run Algorithm~\ref{alg:k_update};
    \EndFor
\end{algorithmic}
\label{alg:dfbsde}
\end{algorithm}

%% file: penalty_algorithm.tex
\begin{algorithm}[t]
\caption{Penalty Function Update}
\begin{algorithmic}[1]
    \State {\bf Given}: $\mathbf{k}$: Boundary steepness, $\delta$: Change of boundary steepness, $\beta$: Update threshold, $\gamma$: Threshold change ratio, $n_I$: Iteration number, $\Delta$: Boundary steepness change acceleration, $\eta$: Update interval, $\eta'$: Max interval, $\Delta_\delta$: Threshold change acceleration;
    \If{state trajectory not inside constraint boundary}
        \If{$(n_I\ \mathbf{mod}\ \eta) = 0$}
            \State Calculate $\sigma_\mathbf{C}^2$ = variance of the costs $\{\mathbf{C}_1, \cdots, \mathbf{C}_\eta\}$ for the past $\eta$ episodes;
            \If{$\sigma_\mathbf{C} < \beta$ or $(n_I\ \mathbf{mod}\ \eta^\prime) = 0$}
                \State $\mathbf{k} = \mathbf{k} + \delta$, $\delta = \delta + \Delta_\delta$, $\beta = \gamma\beta$, $\gamma = \gamma + \Delta$;
            \EndIf
        \EndIf
        \If{$\delta < 0$}
            \State $\delta = 0$;
        \EndIf
        \If{$\gamma > 1$}
            \State $\gamma = 1$;
        \EndIf
    \Else
        \State Update penalty function with new parameters.
    \EndIf
\end{algorithmic}
\label{alg:k_update}
\end{algorithm}

%% file: simulation.tex
\section{Simulations}
\label{sec:simulations}
In this section, we show the performance of our control algorithm using a continuous nonlinear system: the cart-pole for a swing-up task; and a hybrid nonlinear system: a five-link biped for walking. The trained model was evaluated over 256 trials for all systems. For the cart-pole experiments, two LSTM layers with size 16 are used at each time step, followed by a dense layer with an output size corresponding to the state dimension. For fixed initial states, the initial value function estimation uses a trainable weight of size one. The initial hidden state and cell state for the LSTM layers are estimated using a trainable weight of size 16. For the five-link biped experiments, we used two LSTM layers with size 32, followed by a dense layer with size 10. The initial value function network has four layers with size $[8,\ 16,\ 8,\ 1]$. The initial hidden state network consists of four different networks for estimating the initial hidden state and the initial cell state of the two LSTM layers; they all consist of two dense layers that have an output size of 8.

\subsection{Cart-pole Swing-Up Task I}
\label{sec:cartpole1}
\begin{figure*}[t!]
    \centering
    \includegraphics[width=\textwidth]{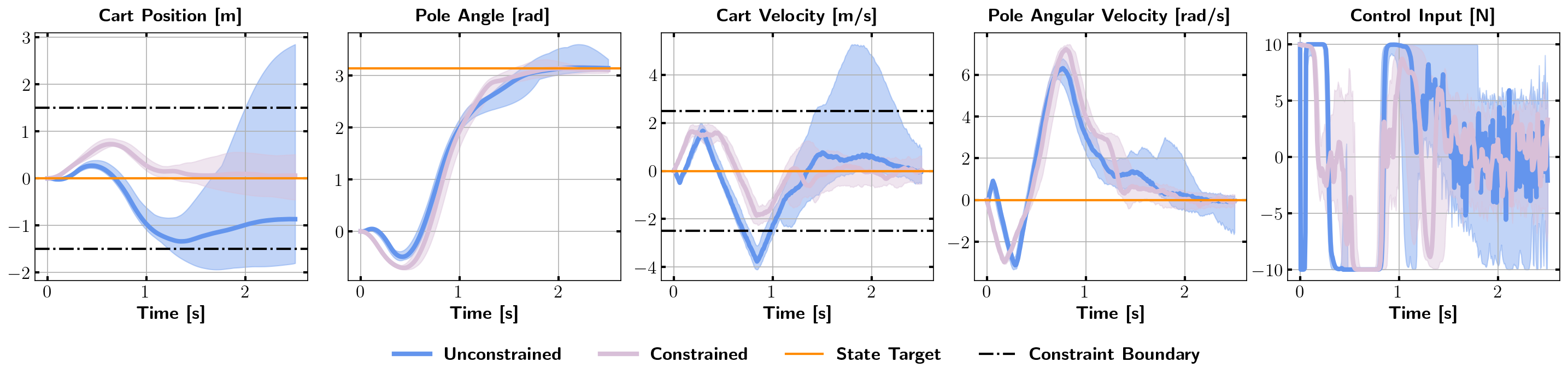}
    \caption{This figure compares the performance between the state constrained (purple) and unconstrained (blue) controller. The demonstrated performance is from 256 trials. The darker blue and purple curves show a sampled trajectory from each setting. The light blue and purple regions show the state space each controller covers. The black dashed lines are the state constraints, and the orange lines give the state targets.}
    \label{fig:comparison}
\end{figure*}

\begin{figure*}[t!]
    \centering
    \includegraphics[width=\textwidth]{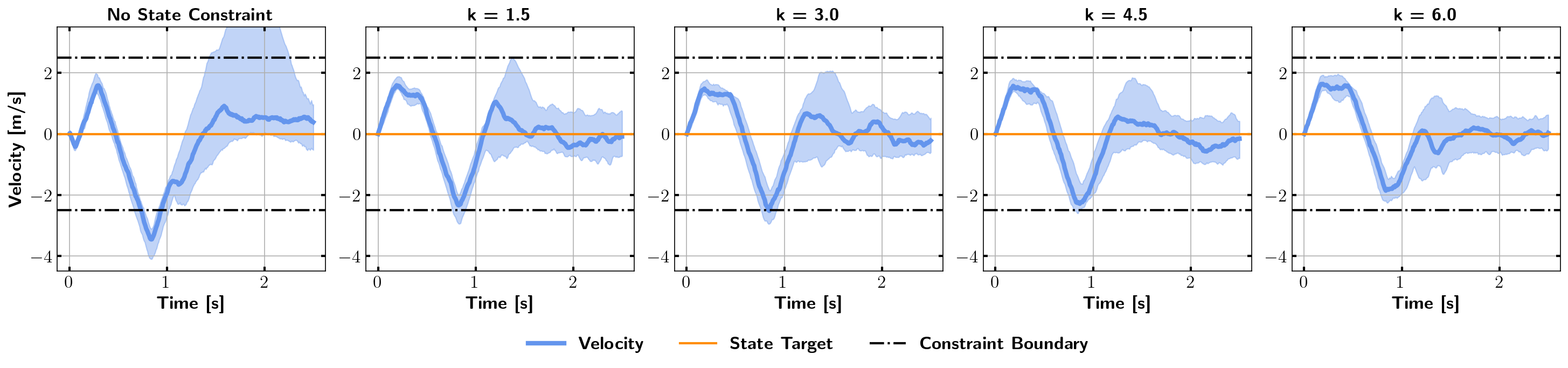}
    \caption{The color scheme follows Figure~\ref{fig:comparison}. The leftmost figure shows the trajectory generated by the unconstrained controller violates the constraints. By applying the penalty function and following the adaptive update scheme, the velocity trajectory retreats inside the constraint boundaries and eventually satisfies the constraints.}
    \label{fig:k_values}
\end{figure*}

The task for the cart-pole system is to swing the pole up and stabilize it at the top. The cart-pole dynamics are
\begin{align}
    (M+m)\Ddot{x} - m\ell\sin\theta\Dot{\theta}^2 + m\ell\cos\theta\Ddot{\theta} &= u\\
    m\ell^2\Ddot{\theta} + m\ell\cos\theta\Ddot{x} + mg\ell\sin\theta &= 0.
\end{align}
The cart position and pendulum angle are represented by $x$ and $\theta$, respectively. The pendulum angle is zero when it is pointed downwards. The state of the cart-pole system is $[x,\ \theta,\ \Dot{x},\ \Dot{\theta}]^T$. In our experiments, we set the cart mass to $M = 1.0$kg, the pole mass to $m = 0.01$kg (point mass at the tip), the pole length to $\ell = 0.5$m, $\x_0 = \mathbf{0}_{4\times1}$, and the target state as $[0, \pi, 0, 0]^T$. The control is saturated at $\pm10$N, the cart position is constrained between $\pm1.5$m, and the cart velocity is constrained between $\pm2.5$m. The time horizon is chosen to be $2.5$ sec and the time step is $\Delta{t} = 1/110$ sec (this specific value of $\Delta{t}$ is chosen randomly). We use the state cost function
\begin{figure}[t!]
    \centering
    \includegraphics[width=0.3\textwidth]{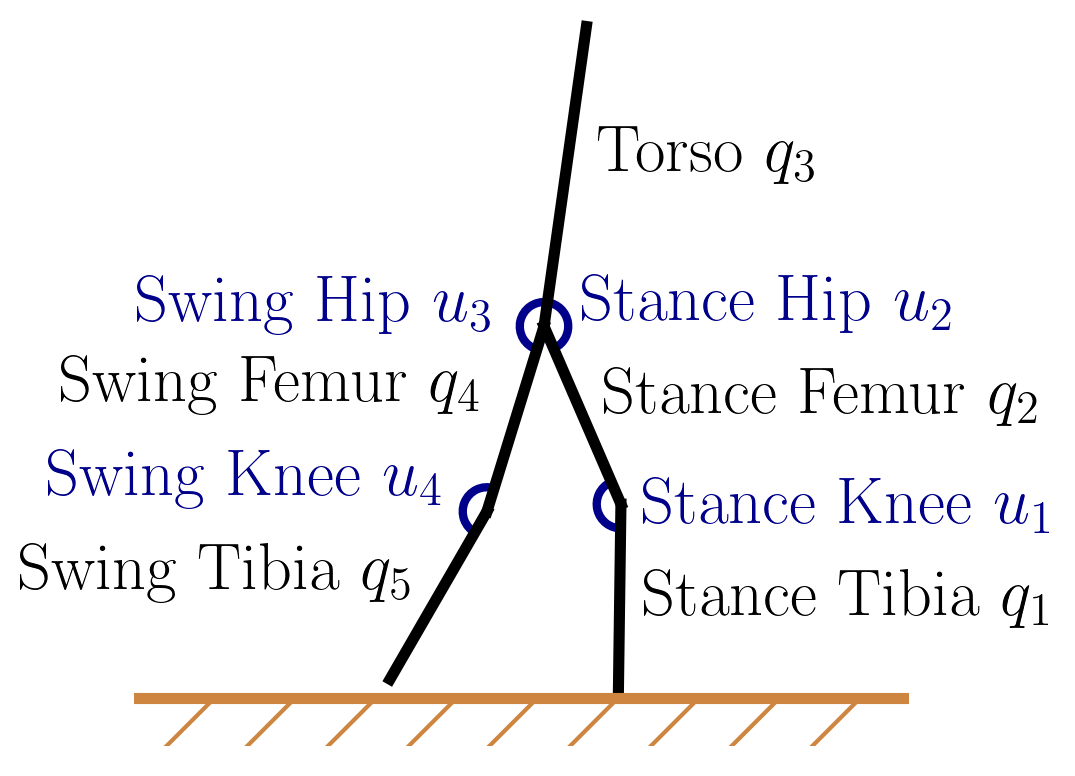}
    \caption{Illustration of five-link biped model.}
    \label{fig:modelIllustration}
\end{figure}
\begingroup
\allowdisplaybreaks
\begin{subequations}
\label{eq:exp_cost}
\begin{align}
    \bar{\mathbf{q}}(\x) &= \frac{1}{2}(\x - \bar{\x})^T\mathbf{Q}(\x - \bar{\x}) + \mathbf{p}(\x)\\
    \mathbf{q}_N(\x) &= \frac{1}{2}(\x - \bar{\x})^T\mathbf{Q}_N(\x - \bar{\x})
\end{align}
\end{subequations}
\endgroup
where $\bar{\x}$ is the target state, $\mathbf{Q}$ is the cost weight matrix for the state, and $\mathbf{Q}_N$ is the terminal state cost matrix. We use the control cost defined in~\eqref{eq:control_constraint_S}. The cost matrices are chosen as $\mathbf{Q} = \mathbf{Q}_N = \mathrm{diag}([0.5,\ 1.0,\ 0.1,\ 0.1])$ and $\mathbf{R} = [0.1]$. $\mathbf{p}(\cdot)$ is the penalty function, in this example $\mathbf{p}(\cdot)$ is a logistics function-based penalty function~\eqref{eq:soft_constraint}, with $\mathbf{c}(\x) = [x, \dot{x}]^T$ and $\mathbf{b}_{\max} = -\mathbf{b}_{\min} = [1.5, 2.5]^T$. Fig.~\ref{fig:comparison} shows the state trajectories generated by the trained controller. The state trajectory generated by the trained constrained DFBSDE controller satisfies the state constraints, while they are violated significantly under the trained unconstrained controller. The effect of the penalty function can be immediately seen in Fig.~\ref{fig:k_values}, even with a small $k$ value. With $k = 1.5$, the part of the trajectory that lies outside the constraint boundaries is greatly reduced. After gradually increasing $k$ to $6.0$, following Algorithm~\ref{alg:k_update}, the entire cart velocity trajectory lies within the constraint boundaries. We also tested directly training with $k = 6.0$ without the adaptive update. The algorithm becomes numerically unstable after one epoch due to large gradients. This demonstrates that our method provides a stable training scheme.

\subsection{Five-Link Biped Walking Task}
The biped model of interest is shown in Fig.~\ref{fig:modelIllustration}, derived from~\cite{1234651}. The leg on the ground is the stance leg, the leg in the air is the swing leg, and the link representing the upper body is the torso. The biped state $\mathbf{x}\in\mathbb{R}^{10\times1}$ consists of the link angles with respect to the horizontal plane $\mathbf{q}\in\mathbb{R}^{5\times1}$ and the link angular velocities $\dot{\mathbf{q}}\in\mathbb{R}^{5\times1}$, i.e., $\mathbf{x} = [\mathbf{q}^T,\ \dot{\mathbf{q}}^T]^T$. Note that the $\mathbf{q}$ denoted here is different from the state cost function $\mathbf{q}(\x)$. The biped model has four actuated joints at the knee and hip joints. The control $\mathbf{u}\in\mathbb{R}^{5\times1}$ consists of the applied torque at the actuated joints $\mathbf{u} = [0,\ u_1,\ u_2,\ u_3,\ u_4]^T$, where the $0$ corresponds to the unactuated stance ankle. The output of the controller will only consist of $\Tilde{\mathbf{u}} = [u_1,\ u_2,\ u_3,\ u_4]$, then a mapping matrix $T = [\mathbf{0},\ \mathbf{I}_{4\times4}]$ is used to obtain $\mathbf{u} = T\Tilde{\mathbf{u}}$. The parameters of the biped model can be found in~\cite{DBLP:journals/siamrev/Kelly17}. The five-link biped dynamics is
\begin{equation}
    \mathbf{M}(\mathbf{q})\ddot{\mathbf{q}} + \mathbf{C}(\mathbf{q}, \dot{\mathbf{q}})\dot{\mathbf{q}} + G(\mathbf{q}) = \mathbf{u},
\end{equation}
where $\mathbf{M}(\mathbf{q})\in\mathbb{R}^{5\times5}$ is the inertia matrix, $\mathbf{C}(\mathbf{q}, \dot{\mathbf{q}})\in\mathbb{R}^{5\times5}$ the Coriolis matrix, and $G(\mathbf{q})\in\mathbb{R}^{5\times1}$ the gravitational terms. The dynamics is in the form of~\eqref{eq:SDE2} after considering stochastic noise. A heel strike (HS) occurs when the biped comes into contact with the ground, which signals the termination of the current step, and initiation of the next step. During this transition, the swing and stance legs are swapped. We assume this transition is instantaneous, the biped is symmetric, and the new swing leg leaves the ground once the HS occurs (i.e., no double support phase). The joint indexing depends on the current swing and stance leg definition. Defining the joint angles right before HS as $\mathbf{q}^-\in\mathbb{R}^{5\times1}$ and the joint angles right after HS as $\mathbf{q}^+\in\mathbb{R}^{5\times1}$, we have $\mathbf{q}^+ = \tilde{\mathbf{I}}_{5\times5}\mathbf{q}^-$, where $\tilde{\mathbf{I}}_{5\times5}\in\mathbb{R}^{5\times 5}$ is the anti-diagonal matrix. The HS creates an instantaneous change in angular velocity. Defining the angular velocity before HS as $\dot{\mathbf{q}}^-\in\mathbb{R}^{5\times1}$ and the angular velocity after HS as $\dot{\mathbf{q}}^+\in\mathbb{R}^{5\times1}$, we have
$\mathbf{x}^{+} = \mathbf{f}_H(\mathbf{x}^{-})$, with $\mathbf{f}_H:\mathbb{R}^{10\times1}\rightarrow\mathbb{R}^{10\times1}$ being the deterministic HS transition map~\cite{DBLP:journals/siamrev/Kelly17}.

\begin{figure}[t!]
    \centering
    \includegraphics[width=0.5\textwidth]{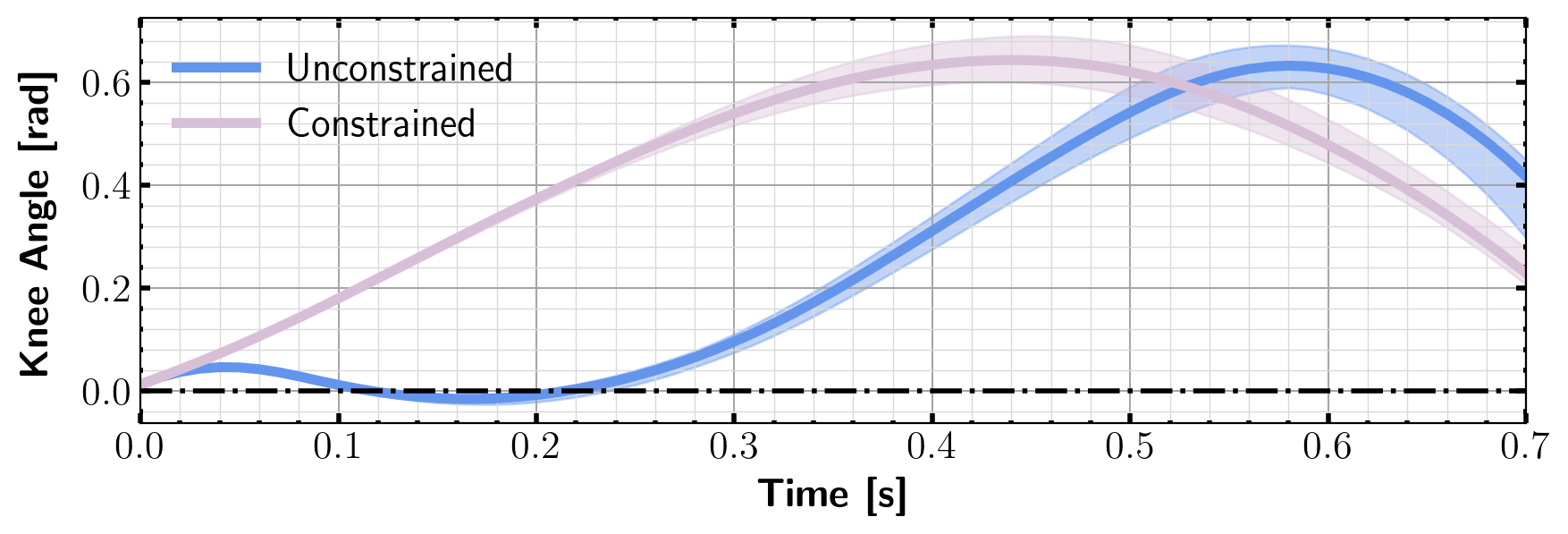}
    \caption{Comparison of knee angles for constrained and unconstrained systems. To avoid hyperextending, the knee angle should be positive, i.e., above the black dashed line.}
    \label{fig:knee_angles_comparison}
\end{figure}

\begin{figure}[t!]
    \centering
    \includegraphics[width=0.49\textwidth]{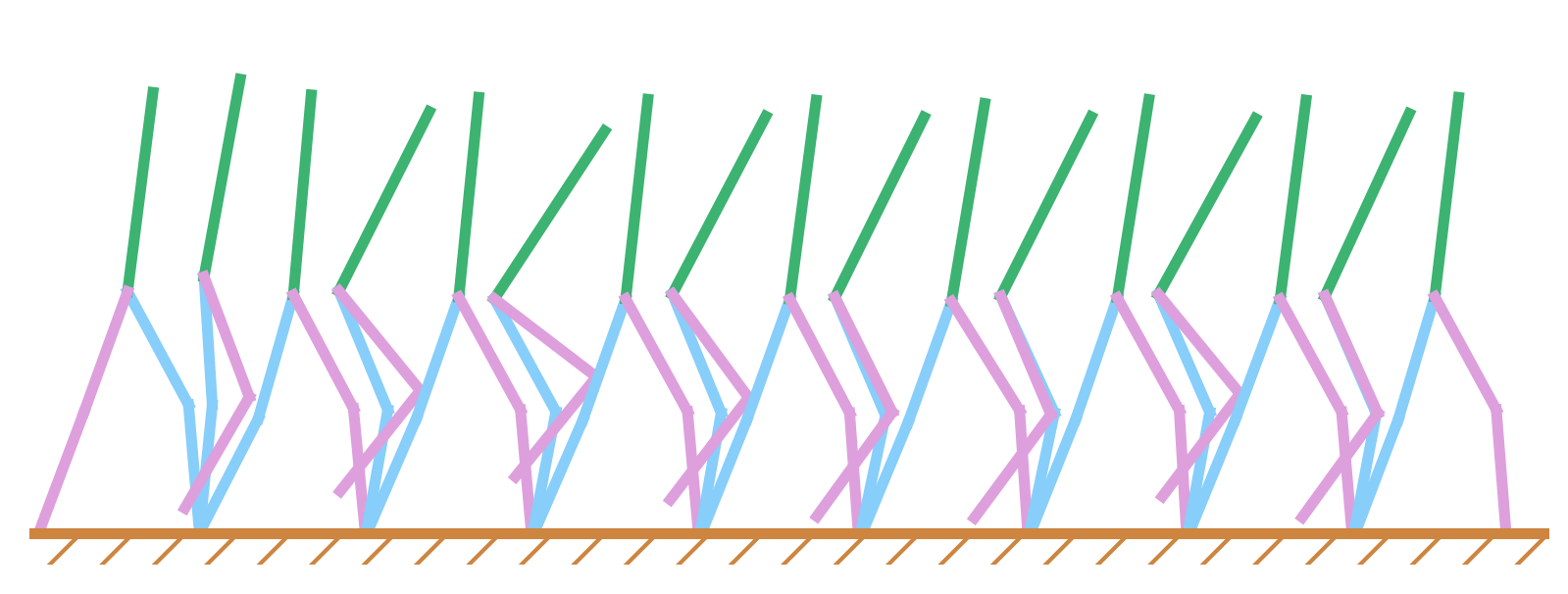}
    \caption{In this figure, the motion generated by our proposed controller ensemble is shown. The torso is depicted in green, the swing leg in purple, and the stance leg in blue.}
    \label{fig:multi_step}
\end{figure}

To generate realistic walking motion, we need to avoid hyperextension in the knee. We transform this constraint into a variant of the ReLU-based penalty function in~\eqref{eq:penalty_relu}: $\mathbf{p}(\mathbf{x}) = \alpha\mathbf{ReLU}(-(q_4 - q_5))$ where $\alpha$ is a weighting coefficient for the penalty function, and $q_4$ and $q_5$ are defined in Fig.~\ref{fig:modelIllustration}. A quadratic cost is applied to the control. We choose $\mathbf{R} = \mathrm{diag}([2,\ 0.2,\ 0.2,\ 2])$. This choice of $\mathbf{R}$ implies that hip motors are more powerful than knee motors, which is true for many robots. The state cost $\mathbf{q}(\mathbf{x})$ is the weighted distance between the state $\mathbf{x}$ and a nominal terminal state $\bar{\mathbf{x}}$ plus the penalty
\begin{equation}
\label{eq:biped_state_cost}
    \bar{\mathbf{q}}(\mathbf{x}) = 1/2(\mathbf{x} - \bar{\mathbf{x}})^T\mathbf{Q}(\mathbf{x} - \bar{\mathbf{x}}) + \mathbf{p}(\x)
\end{equation}
with $\mathbf{Q} = 10\mathbf{I}_{10\times10}$. The terminal state cost has the same form as~\eqref{eq:biped_state_cost}, the sole difference is replacing the state cost matrix with $\mathbf{Q}_N = 10\mathbf{Q}$. The target state is chosen to be
\begin{align}
    \bar{\mathbf{x}} =& \Big[0.10, 0.50, -0.10, -0.35, -0.40, -1.50\nonumber\\
                      & -0.50, 0.00, -0.55, -2.00\Big]^T.
\end{align}
A comparison between the motion generated by the constrained and unconstrained DFBSDE controller is given in Fig.~\ref{fig:knee_angles_comparison}. The unconstrained controller hyperextends the swing knee, while the constrained controller does not. Since the chosen penalty function could constrain the system directly, the $k$ value doesn't need to be updated. In previous examples, the penalty functions are logistic function based. This works well when the constraint set is large since it creates a buffer between the interior of the constraint set and its boundary. However, the constraint set for locomotion tasks is relatively small, making ReLU functions a better candidate. For the five-link biped experiments, we use a time step of 0.01s.

Variations in the initial states make dissecting a multi-step walking problem into multiple single-step walking problems a challenging task. We train a robust controller to deal with this issue. We find it difficult for a single controller to handle all possible initial configurations. Thus, we use an ensemble of controllers, where each controller handles a range of initial configurations around a nominal state. For the duration of one footstep, only one controller is used, which is the controller in the ensemble that has the shortest distance between the initial state and its nominal state. The nominal state of a controller is known a priori. After training, the motion generated by this approach is shown in Fig.~\ref{fig:multi_step}, where the ensemble size is three. The corresponding swing knee angle is shown in Fig.~\ref{fig:knee_angles_comparison}. It can be seen that no hyperextension occurred, which shows the effectiveness of enforcing the state constraints. On average, the initial range each controller can handle spans 6.03 deg for the joint angles and 19.01 deg/sec for the joint velocities. Fig.~\ref{fig:multi_step} shows a tucking motion generated by the torso and the swing leg. This is due to our biped model's relatively small control authority over the torso. Without the forward angular momentum generated by the tucking motion, the torso gradually tilts back over multiple steps. Since there is no direct control over the torso, it is difficult for the DFBSDE controller to recover. We also compared the computation time of our proposed method for obtaining the control for one time step with trajectory optimization~\cite{DBLP:journals/siamrev/Kelly17} (TO). As expected, our approach generates comparable results while being computationally more efficient. Solving for the control action requires 0.96s for a Hermite-Simpson TO approach, compared with 5.6ms for our proposed method.

\begin{figure}[t!]
    \centering
    \includegraphics[width=0.49\textwidth]{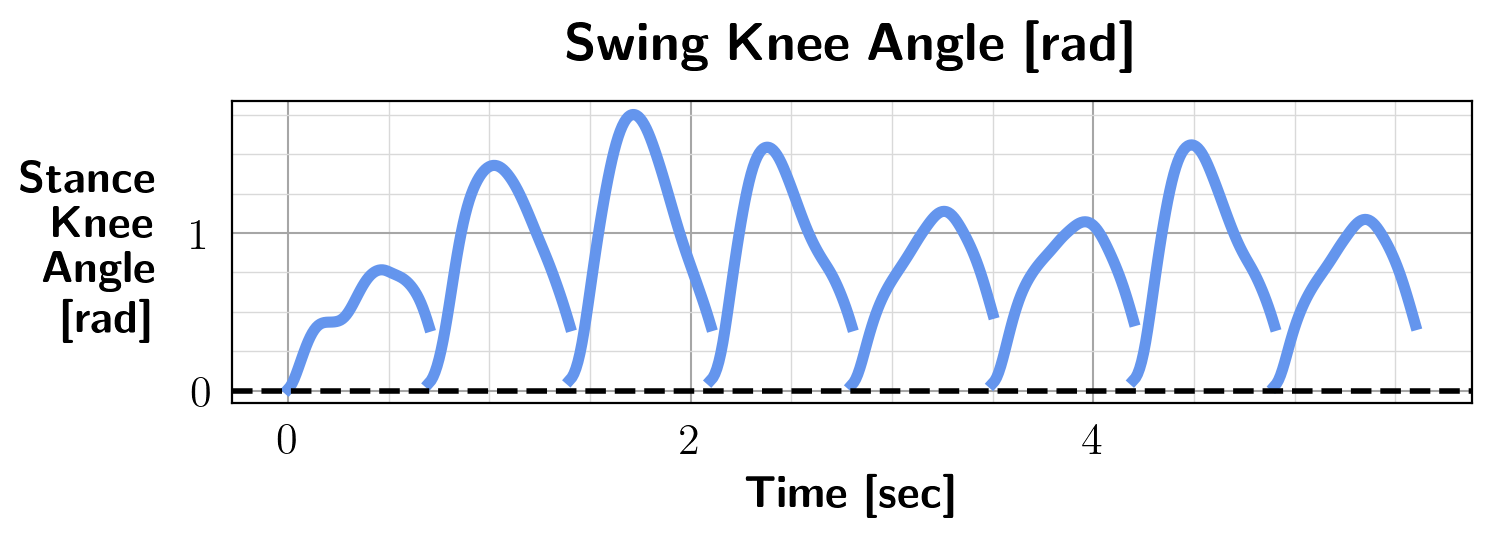}
    \caption{This figure shows the swing knee angles for the motion depicted in Fig.~\ref{fig:multi_step}.}
    \label{fig:multi_step_controls}
\end{figure}

%% file: conclusion.tex
\section{Conclusion}
In this paper, we solved SOC problems with state constraints using an LSTM-based DFBSDE framework which alleviates the curse of dimensionality and numerical integration issues.  The state constraints are applied using an adaptive update scheme, significantly improving training stability. We also show how to adapt the algorithm to handle HD systems in addition to the continuous dynamics setting. The efficacy of our approach is demonstrated on a cart-pole system and a five-link biped which has hybrid dynamics. 